\documentclass[11pt,twoside]{article}
 
\usepackage{asp2006}
\usepackage{epsf}
\usepackage{psfig}
\usepackage{lscape}
 
\begin{document}

\title{An Atlas of GALEX UV Images of
Interacting Galaxies}
\author{
Mark L. Giroux$^1$,
Beverly J. Smith$^1$,
Curtis Struck$^2$,
Mark Hancock$^{3}$, and
Sabrina Hurlock$^1$}
\affil{$^1$East Tennessee
State University, 
$^2$Iowa State University, 
$^3$University of California Riverside}

\begin{abstract} 

We present GALEX ultraviolet images from a survey of
strongly interacting galaxy pairs, and compare with images
at other wavelengths.  The tidal features are particularly
striking in the UV images.  Numerous knots of star formation
are visible throughout the disks and the tails and bridges.
We also identify a possible `Taffy' galaxy in our sample,
which may have been produced by a head-on collision between
two disk galaxies.

\end{abstract}



\section{Introduction}

GALEX imaging 
has shown that some tidal features in interacting galaxies
are quite bright in the UV (\citealp{neff05},
\citealp{hancock07}, \citealp{hancock09}).
In
\citet{smith10a, smith10b}, 
we describe the `Spirals, Bridges, and Tails'
(SB\&T) GALEX imaging survey
of more than three dozen strongly 
interacting galaxies.
Here, we provide an Atlas of some of these
images, and discuss some 
particularly intriguing systems.

\section{Descriptions of Selected Individual Systems}

\noindent{\bf Arp 35:}
Arp 35 (Figure 1) is
a widely separated M51-like system. The brighter northern
galaxy has strong tails
with bright knots of star formation.
The southern galaxy has two
short tails.
The southern tail of the smaller galaxy
is the second bluest tidal feature in FUV $-$ NUV 
in the SB\&T sample.

\noindent{\bf Arp 86:} 
The GALEX images of the M51-like pair Arp 86 
(Figure 1) show a 
clumpy arc of UV emission to the south of the main galaxy, connecting
to the companion.  
This feature is similar to the eastern arc of 
Arp 82 \citep{hancock07}.
This arc lies within the HI disk 
of Arp 86 (\citealp{sengupta09}).
This loop is
also faintly visible in the \citet{arp66} Atlas
image.
The HI map also shows a strong tail to the 
north, looping to the west.
The base of this tail is seen in the GALEX NUV image.
Near the tip of this tail to the northwest, a concentration
of gas with N$_{HI}$ $\sim$ 10$^{20}$ cm$^{-2}$ is 
visible in the HI maps, but is not detected in the 
GALEX images.

\noindent{\bf Arp 100:}
Arp 100 (Figure 2)
is a widely separated elliptical/disk galaxy pair.  The northern
galaxy has two long tails, with four prominent `beads' of
star formation visible
along the brighter northern tail
in the GALEX images.

\begin{figure}
\plottwo{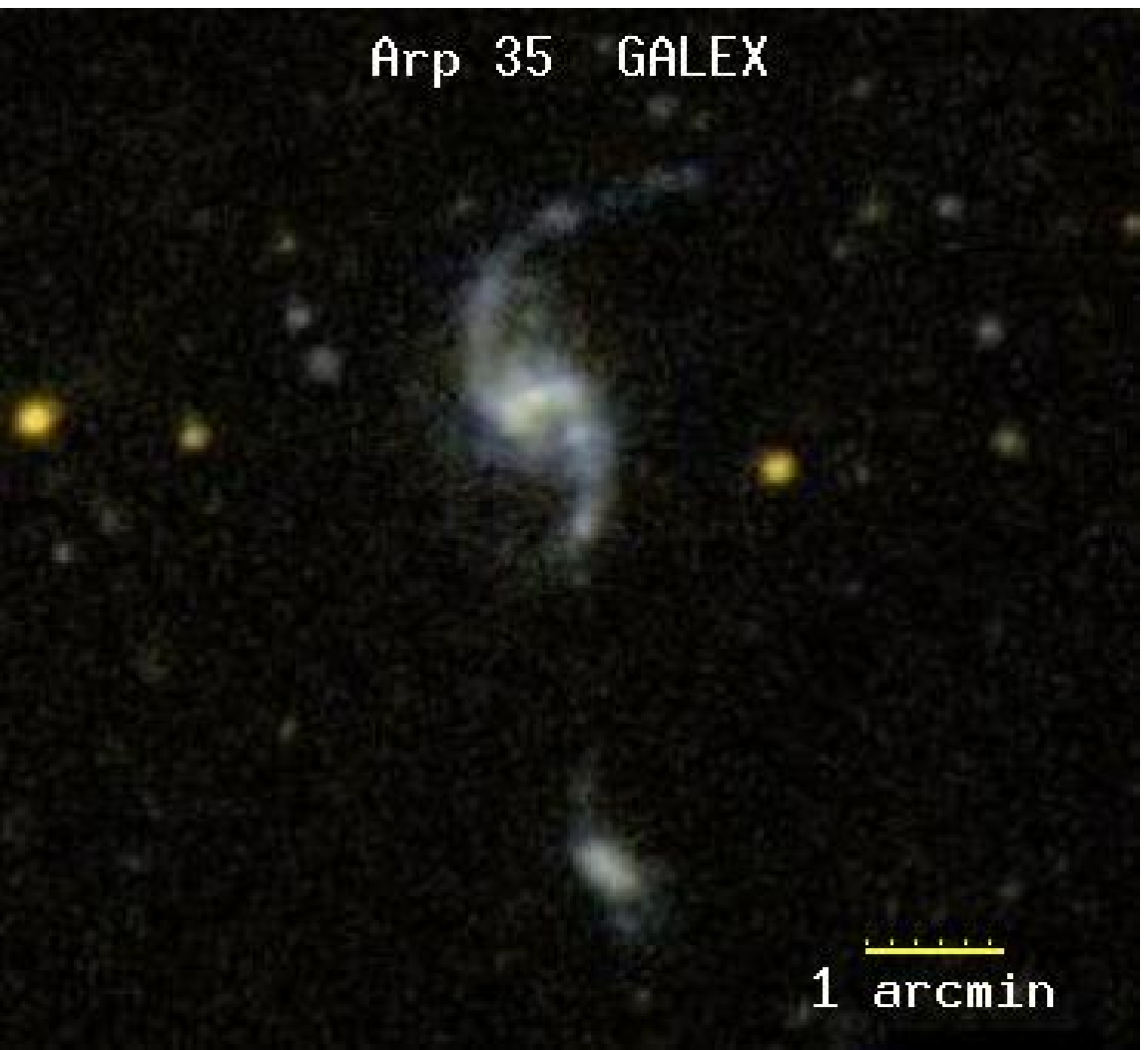}{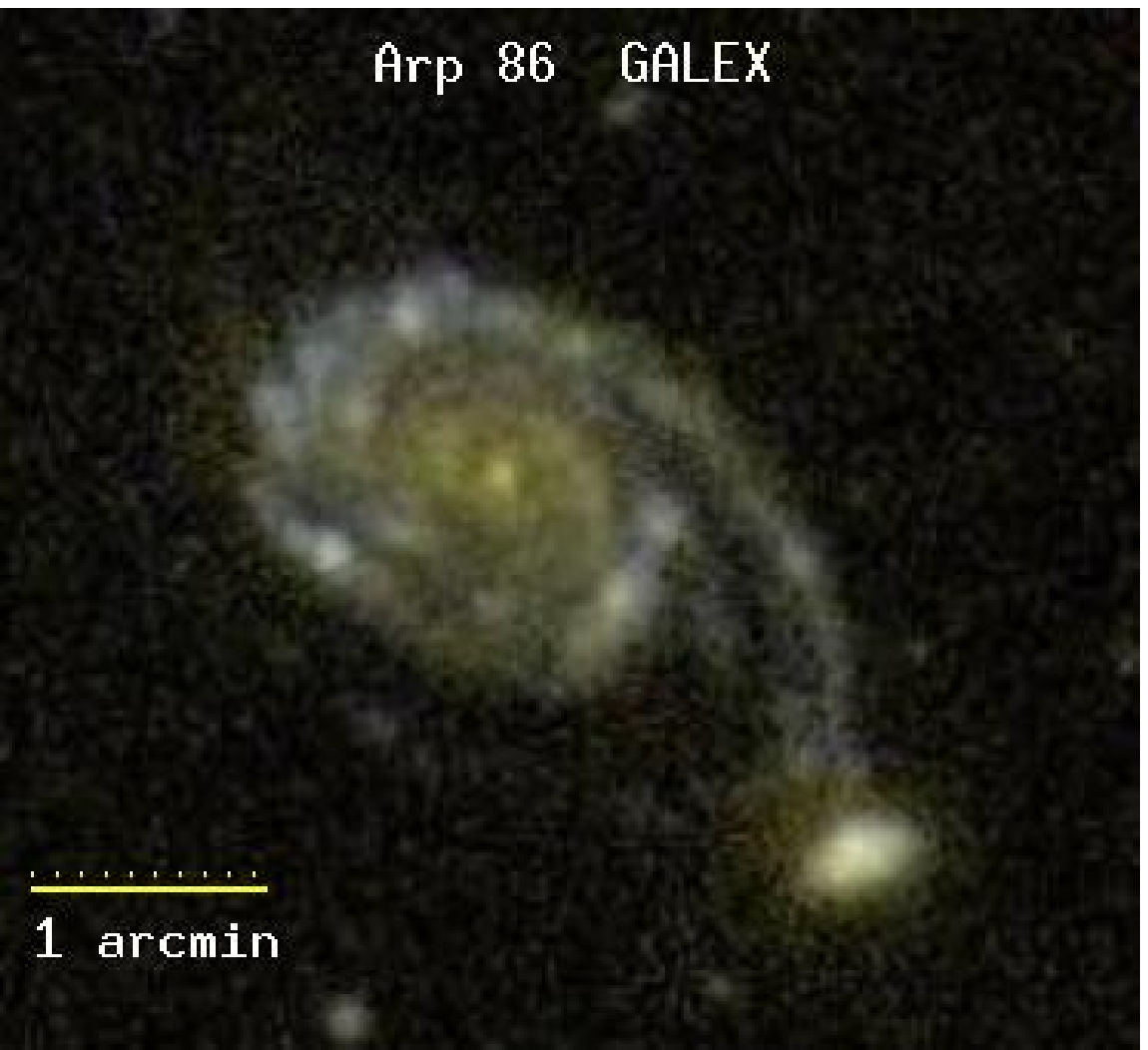}
\caption{
  \small 
GALEX images of Arp 35 (left) and Arp 86 (right).
North is up and east to the left in all figures in this paper.
}
\end{figure}

\begin{figure}
\plottwo{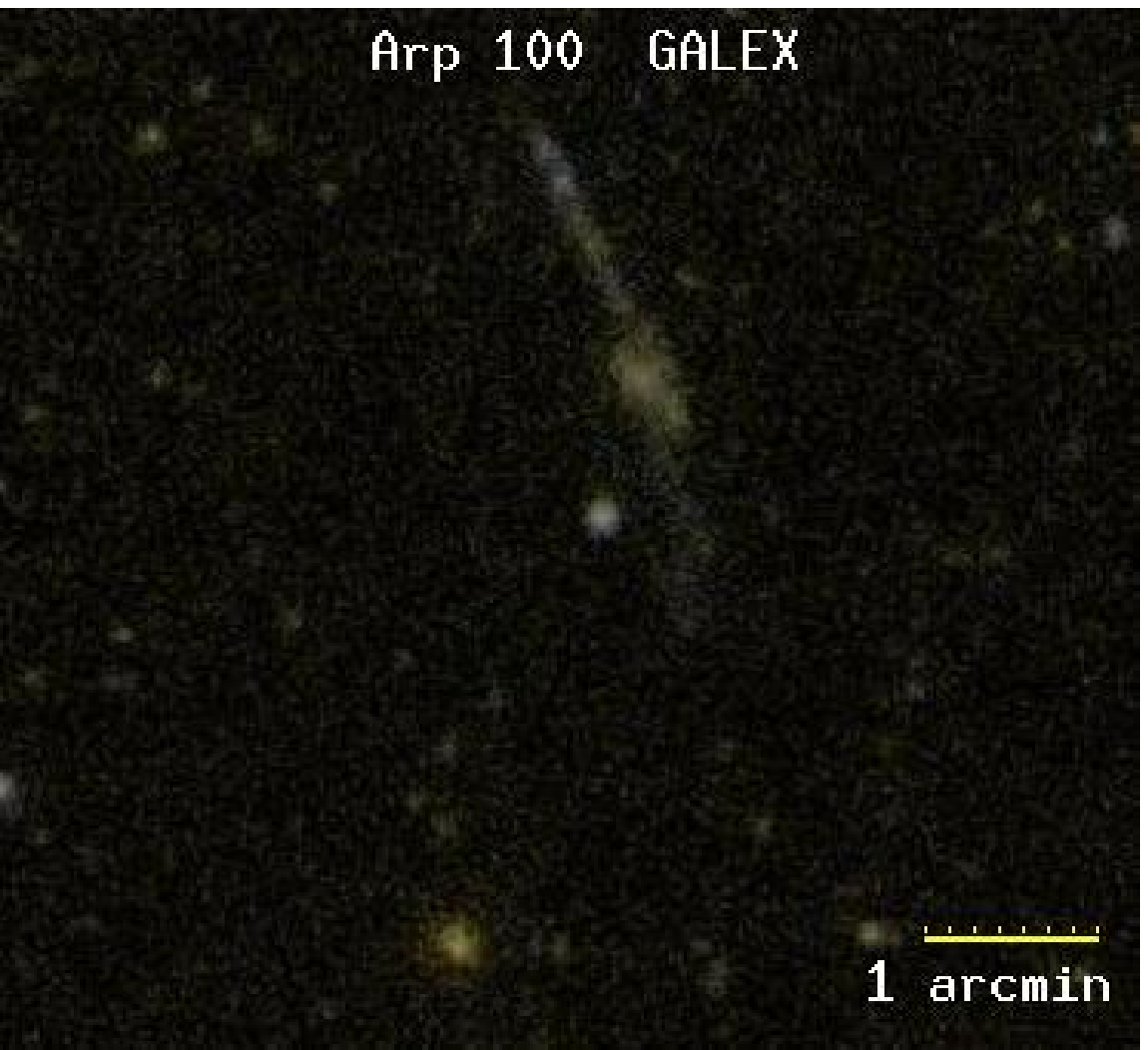}{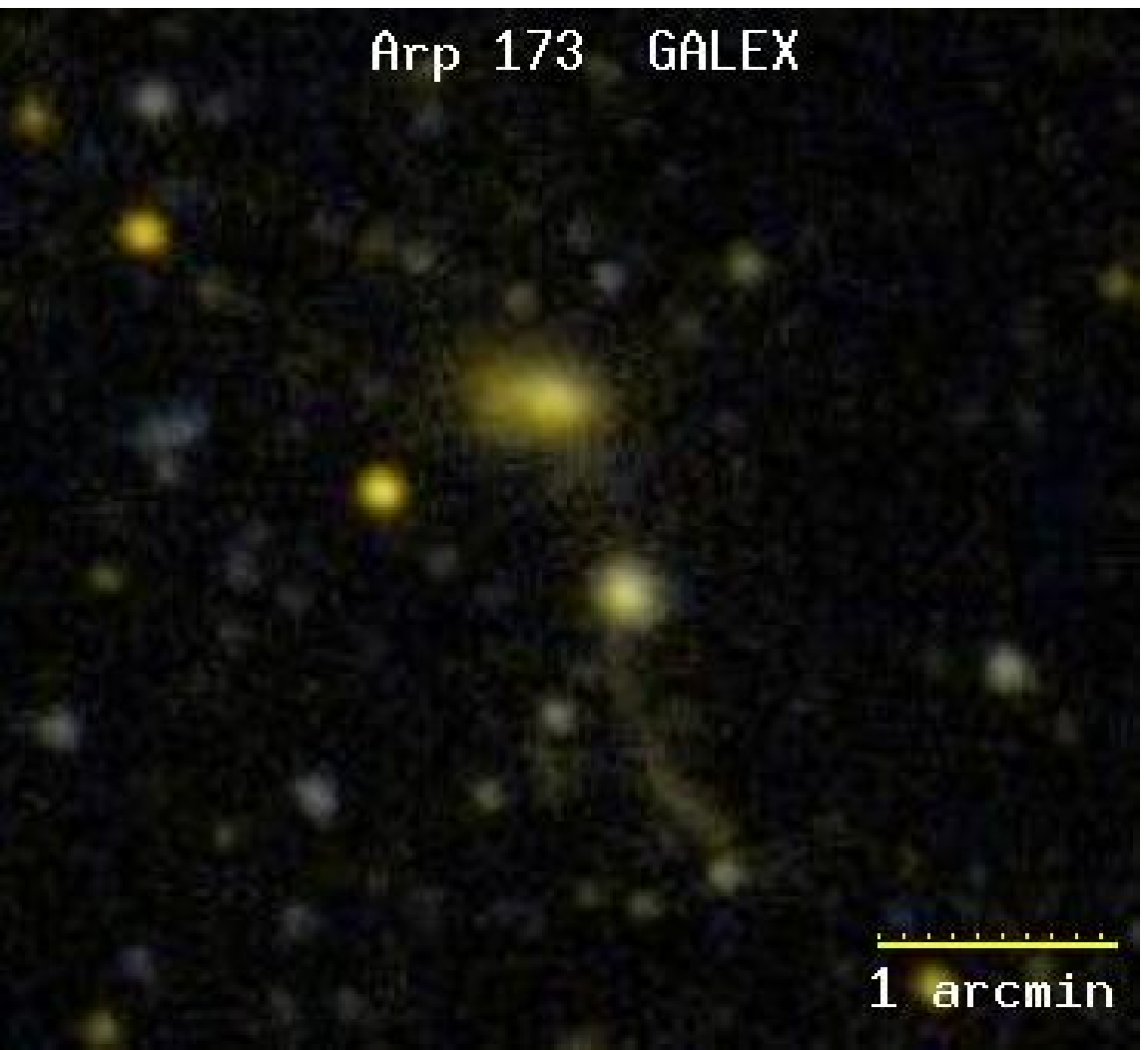}
\caption{
  \small 
GALEX images of Arp 100 (left) and Arp 173 (right).
}
\end{figure}

\begin{figure}
\plottwo{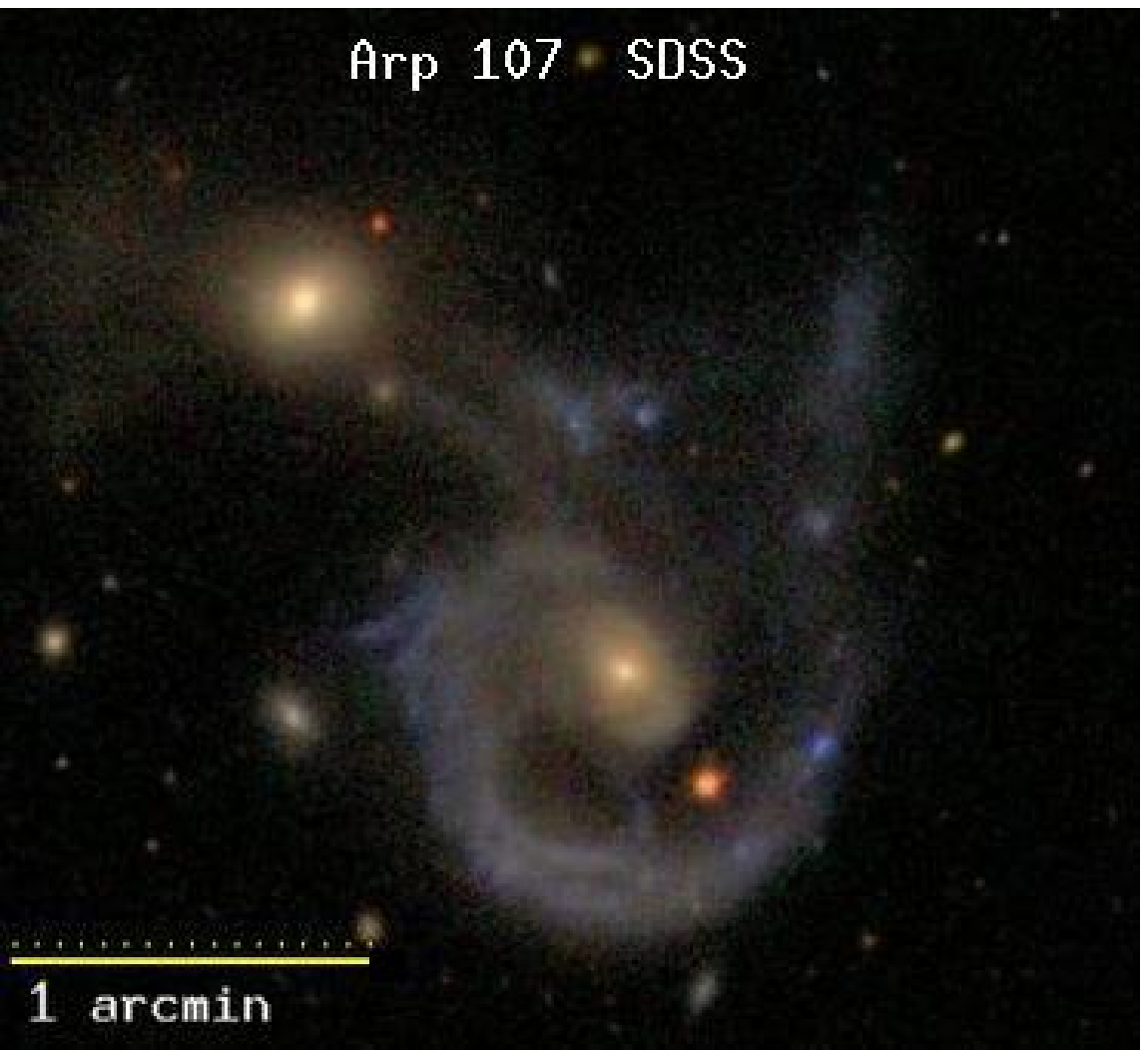}{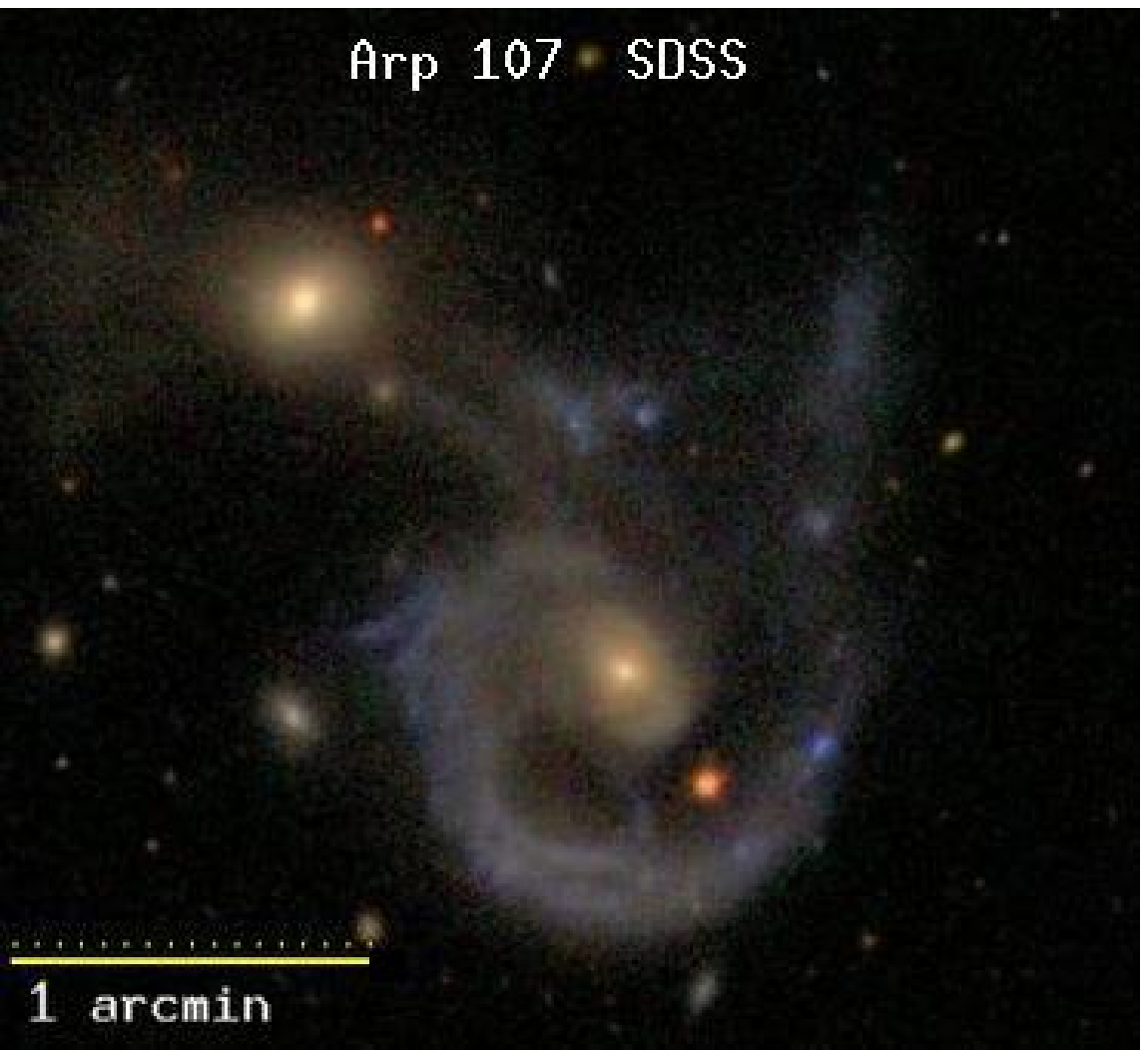}
\caption{
  \small 
The GALEX (left) and SDSS (right) images of Arp 107.
}
\end{figure}

\noindent{\bf Arp 107 (`The Smile'):} 
In the \citet{arp66} Atlas,
the main disk of 
Arp 107 
(Figure 3) shows a prominent spiral arm/partial 
ring-like structure, connected to a short tail.
This galaxy is connected to
the elliptical-like companion 
by a bridge.   
The basic morphology of this system can be produced
by a hybrid model, in between a classical ring galaxy
simulation of a head-on collision
with a small impact parameter, and a prograde planar
encounter \citep{smith05a}.
The ring is quite bright at 8 $\mu$m, 
especially in the south, 
as are two luminous knots of star formation in
the northwestern sector of the system \citep{smith05a}.
The ring and the northwestern
knots of star formation are also very bright in
the GALEX images,
while
the companion is relatively faint.  
In contrast,
the northern tail has a red NUV $-$ g color.

\noindent{\bf Arp 173: }
A long tail is visible to the south of the southern galaxy in Arp 173 (Figure 2).
This has very red UV/optical colors, perhaps in part because its progenitor
is an S0 galaxy with little star formation.

\begin{figure}
\plottwo{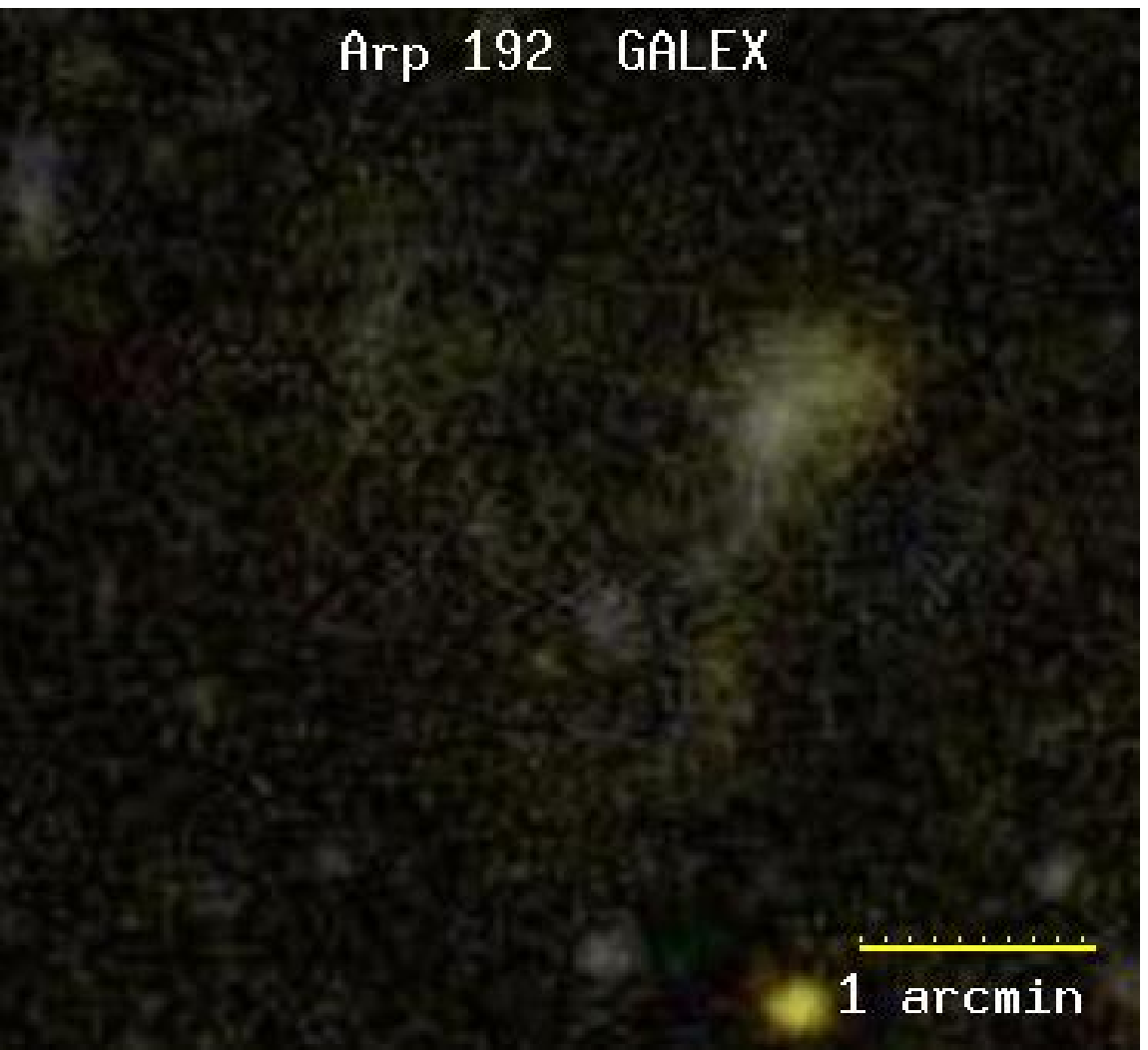}{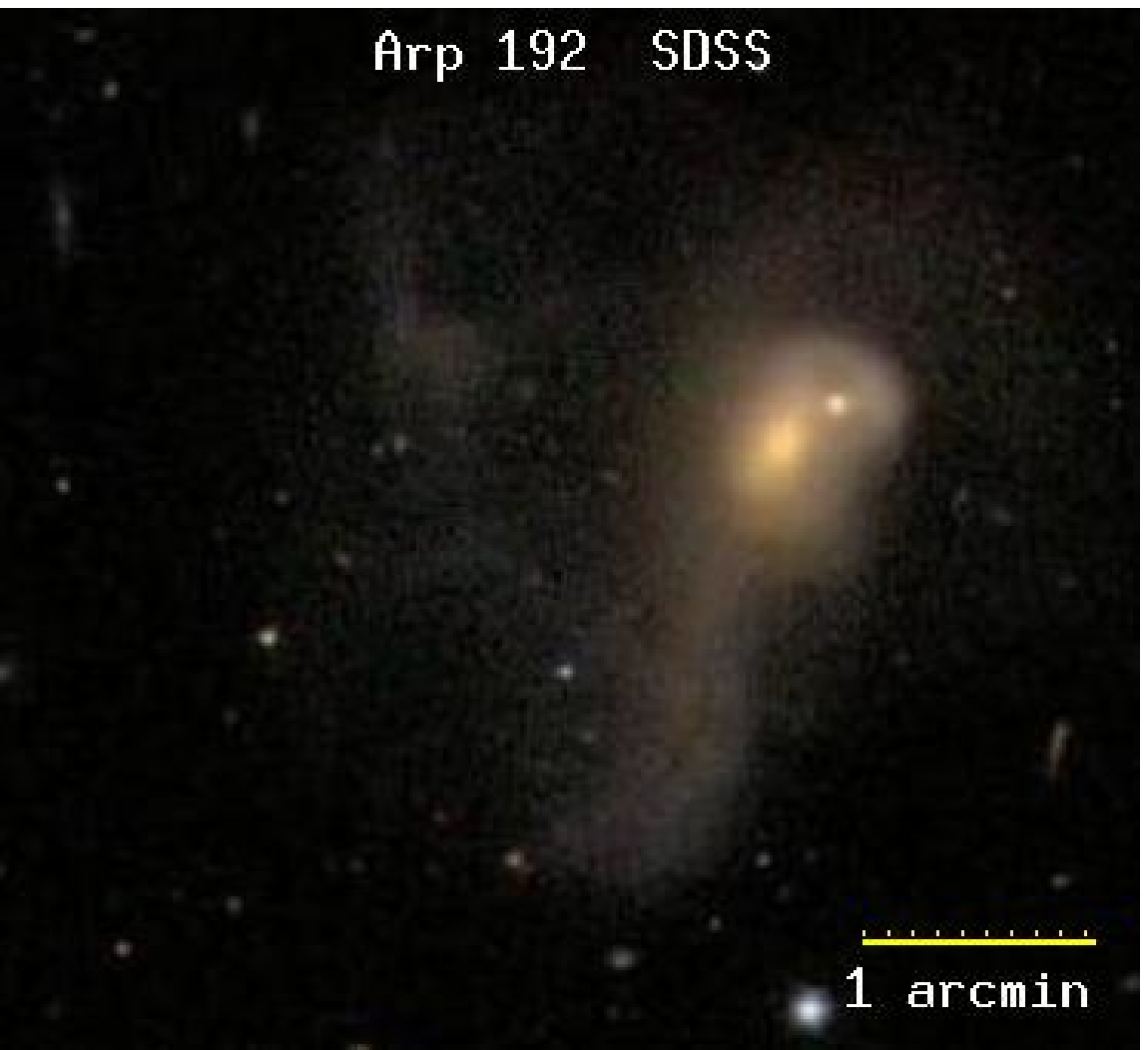}
\caption{
  \small 
The GALEX (left) and SDSS (right) images of Arp 192.
}
\end{figure}

\begin{figure}
\plottwo{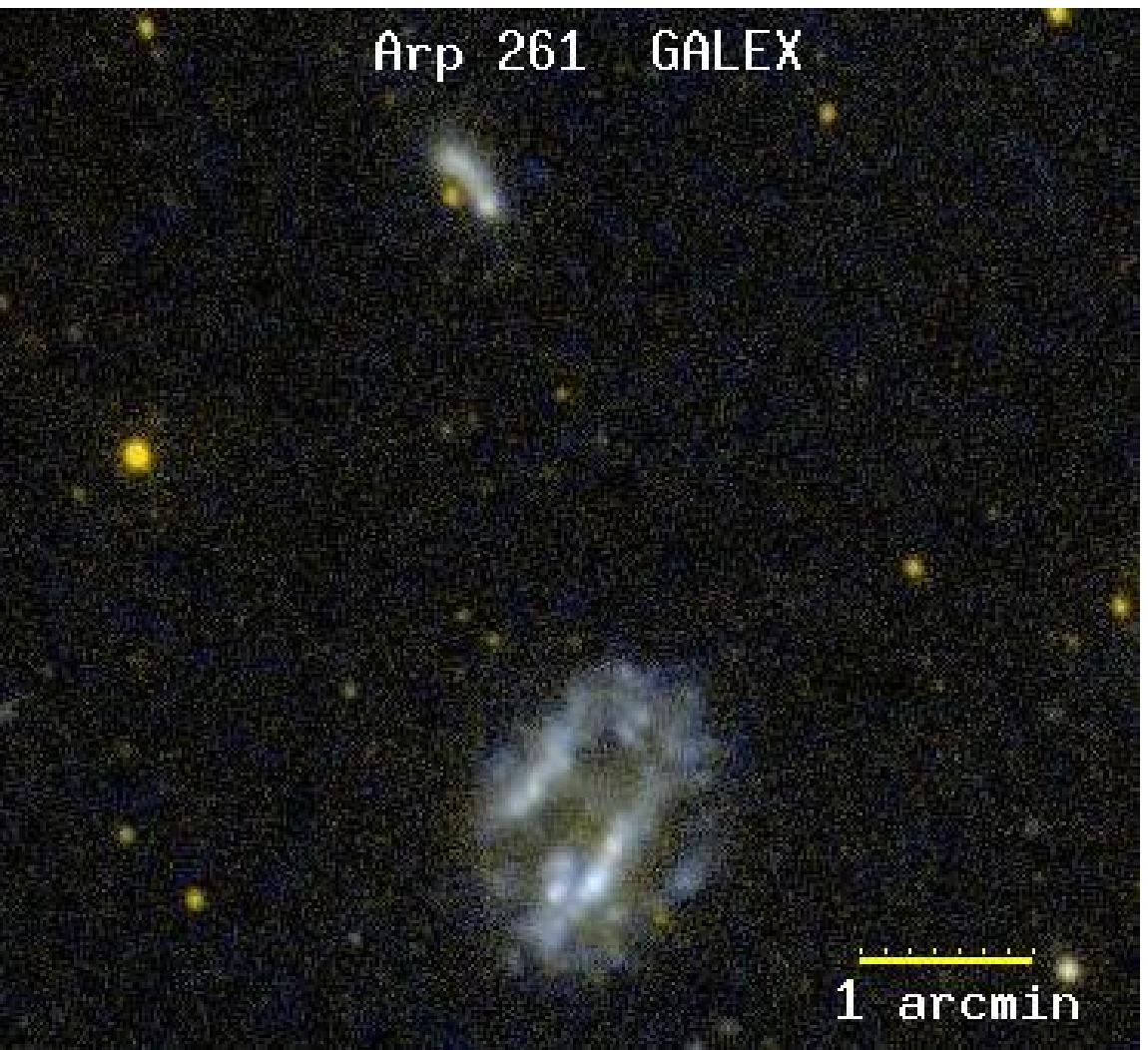}{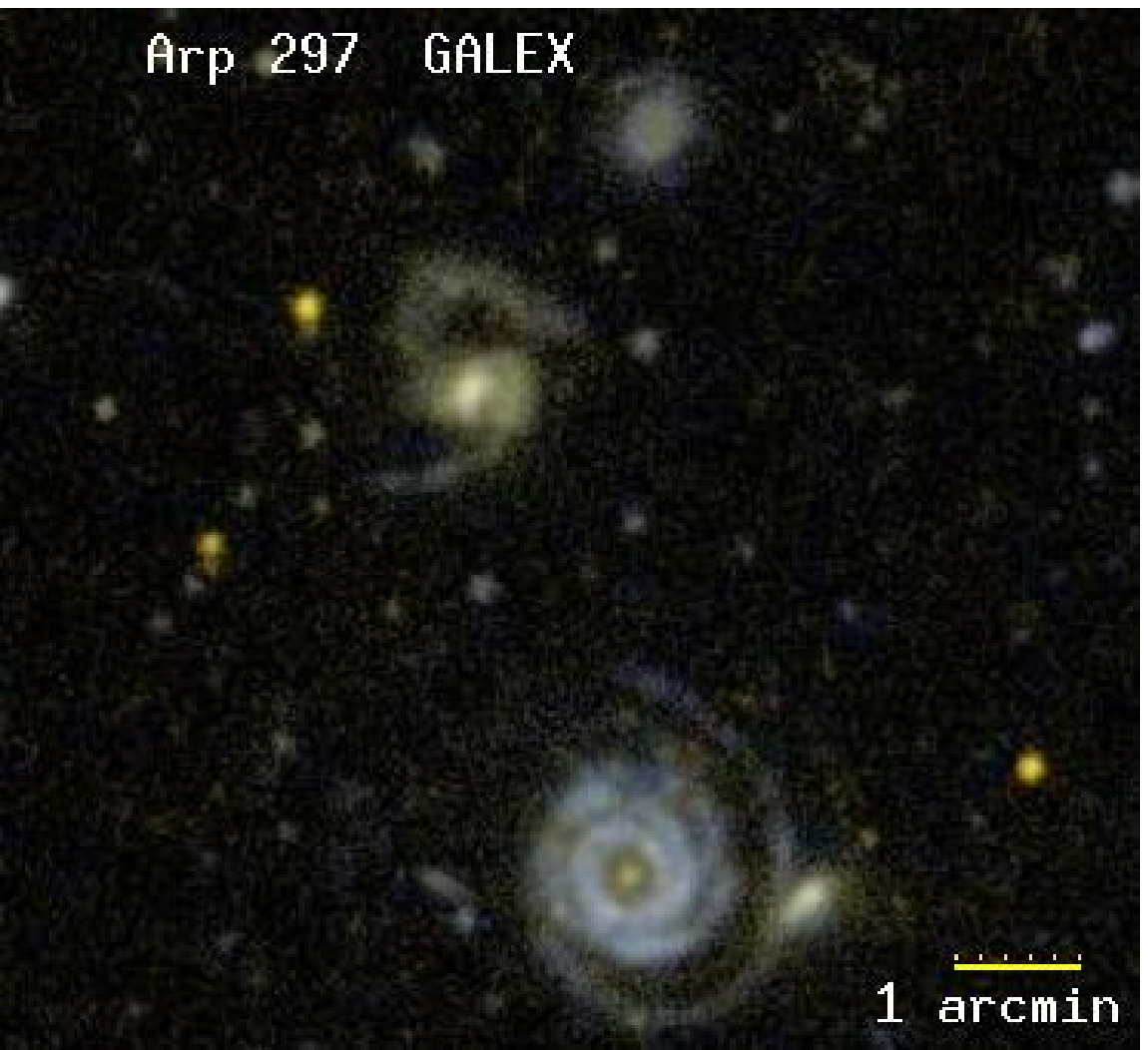}
\caption{
  \small 
GALEX images of Arp 261 (left) and Arp 297 (right).
}
\end{figure}

\noindent{\bf Arp 192:} 
The peculiar system Arp 192 (Figure 4) has a very long distorted tail
that curves to the south and then to the east.   In the SDSS
images, the main body
of this system appears to be a very close pair, with 
a disk galaxy containing a ring galaxy-like
loop and a second compact galaxy.

\noindent{\bf Arp 261:} 
Arp 261 (Figure 5) may be an example of a `Taffy' galaxy,
produced by a direct head-on collision
between two equal-mass gas-rich galaxies.
During such collisions,
the impact may be sufficient to ionize the gas and strip it
from the disks, leaving a large quantity of gas between
the two galaxies \citep{condon93}.  
In Taffy galaxies, the bridge is bright in radio
synchrotron emission due to the magnetic field being stripped
from the galaxy along with the interstellar matter (Condon et al.\ 1993).
Only two Taffy systems
have been identified to date, UGC 813/6 and UGC 12914/5 
\citep{condon93, condon02},
thus the possible existence of a third in the nearby
Universe is of great interest.

Arp 261 is a close pair of edge-on irregular or spiral galaxies, 
with an optical and UV morphology suggesting a recent head-on collision.  
The western galaxy in the pair has a possible ring-like structure,
while an apparent bridge 
is visible to the northeast
between the two galaxies. 
In the NRAO VLA Sky Survey (NVSS), the radio continuum map
shows a bright peak between the two galaxies, suggesting
a possible radio continuum bridge between the
two galaxies, as in UGC 12914/5 and UGC 813/6. 
Arp 261 is much closer than the other two Taffy galaxies,
at only 27 Mpc, and has a lower B luminosity,
suggesting that it may consist of two lower mass dwarf galaxies
rather than spirals.
A third galaxy, KTS 52, to the northeast of the pair, is part of the same system.

\noindent{\bf Arp 297:}
Arp 297 (Figure 5)
consists of two unrelated pairs of galaxies at different redshifts.
The more northern pair includes NGC 5755,
a disturbed
spiral with prominent tails/arms, and
NGC 5753, a more compact galaxy.
In the \citet{arp66} Atlas image, the
southern pair in Arp 297 resembles M51,
with
a face-on grand design spiral
NGC 5754
and
a smaller compact galaxy
NGC 5752.
However,
in the smoothed GALEX and SDSS images
a 6$'$ (110 kpc) long faint tail extending
to the west is visible.
This feature is also
visible
in deep optical
images (P. Appleton, 2005, private communication).

\acknowledgements 

This research was supported by 
NASA LTSA grant NAG5-13079, 
GALEX grant GALEXGI04-0000-0026,
and Spitzer
grants RSA 1353814 and RSA 1379558.

\end{document}